\documentclass[11pt]{elsarticle}   	
\usepackage{amssymb}

\usepackage{lineno,hyperref,geometry,amsmath,appendix}
\modulolinenumbers[5]
\usepackage{bbold}
\newcommand{\gdualn}[1]{\overset{\:{}^{{}^{\boldsymbol{\neg}}}}{\smash[t]{#1}}} 
\def\0{\mbox{\boldmath$\displaystyle\mathbb{O}$}}
\def\C{\mbox{\boldmath$\displaystyle\mathbb{C}$}}
\def\R{\mbox{\boldmath$\displaystyle\mathbb{R}$}}

\def\x{\mbox{\boldmath$\displaystyle\boldsymbol{x}$}}

\def\I{\openone}
\def\s{\mbox{\boldmath$\displaystyle\boldsymbol{\sigma}$}}
\def\p{\mbox{\boldmath$\displaystyle\boldsymbol{p}$}}

\def\openone{\mathbb I}

\journal{Yet to be Submitted}

\begin{document}

\begin{frontmatter}
\title{Theory of spin one half bosons}


\author[mymainaddress]{Dharam Vir Ahluwalia\corref{mycorrespondingauthor}}
\cortext[mycorrespondingauthor]{Corresponding author}
\ead{ahluwalia.icc@charusat.ac.in}


\address[mymainaddress]{International Centre for Cosmology,
P. D. Patel Institute of Applied Sciences,
Charotar University of Science \& Technology, Changa, Gujarat 388421, India}


\begin{abstract}

These are notes on the square root of $4\times4$ identity matrix and  associated quantum fields of spin one half. The method is illustrated by constructing a new mass dimension one bosonic field. 
The locality constraint for the field leads naturally to maximum parity violation. The degrees of freedom carried by the new bosons are different from any massive boson previously encountered and coincide with those carried by spin one half fermions. We thus provide a quantum field suspected to exist by  Dolgov and Smirnov in the context of cosmological neutrinos.
\end{abstract}

\end{frontmatter}
\vspace{21pt}

\noindent
Historically~\cite{Dirac:1928hu}, Dirac equation arose in taking the square root of the dispersion relation
$
p^\mu p_\mu = m^2$
The square root of the left hand side was found to be $\gamma_\mu p^\mu$, where the $\gamma_\mu$ are the celebrated $4\times 4$ matrices of the Dirac framework. The argument naturally leads to 
$(\gamma_\mu p^\mu \pm m \I)\psi(\p) =0$, the Dirac equation in momentum space.
Its solutions, after attending to certain locality phases, later became expansion coefficients of all fermionic matter fields of the standard model~\cite{Weinberg:1995mt}. 
\vspace{5pt}

Modulo the Majorana observation of 1937~\cite{Majorana:1937vz}, there is a general consensus that the Dirac field presents a unique spin one half field that is consistent with Lorentz symmetries and locality.
The uniqueness, however, hinges on the implicit assumption that the square root of a $4\times4$ identity matrix $\I$ multiplying the $m^2$  on the right hand side, is $\I$ itself.  The recent emergence of new spin one-half fermions with mass dimension one provides a strong reason that other roots of $\I$ may lead to new spin one half matter fields, and these may serve the dark matter sector or at the least us provide us a complete set of particle content consistent with basic principles of quantum mechanics and Lorentz symmetries~\cite{Ahluwalia:2019dv}.
\vspace{5pt}

With this background and motivation we here present non-trivial square roots $A$ of $\I$:
$
A A \stackrel{\mathrm{def}}{=} \I.
$
where $A$ is a $4\times 4$ matrix with sixteen complex entries. We classify these roots as antisymmetric and symmetric -- the most general roots  being deferred to our readers . A detailed calculation shows that these depend only on  $A_{13} \stackrel{\mathrm{def}}{=} 
\gamma + i \delta$ and $A_{14} \stackrel{\mathrm{def}}{=} \alpha + i \beta$ with $\alpha,\beta,\gamma,\delta \in \R$. Our results on the roots are collected together in~\ref{AppA}. Unlike the Dirac root $\I$ none of the twenty eight roots we find commute with $\gamma_\mu p^\mu$. This non-commutativity thus requires us to examine the behaviour of the eigenvectors of $A$ under the action of $\gamma_\mu p^\mu$. For Lorentz symmetries not to be violated if $\lambda_i$, $i=1,2,3,4$, are the mentioned eigenspinors
\begin{equation}
A  \lambda_i(\p) = a \lambda_i(\p) , \qquad a \in \R
\end{equation}
then the action of $m^{-1} \gamma_\mu p^\mu$ on $\lambda_i(\p)$, up to a factor of $\pm i$ or $\pm 1$, must be one of the eigenspinors, $\lambda_{j\ne i}(\p)$. 
\vspace{5pt}

\noindent
To illustrate the method I consider the first of the two $A$'s in equation (\ref{eq:a1}). Its eigenspinors, up to constant factors, are
\begin{equation}
\lambda_1 = \left(
\begin{array}{c}
 0 \\
 0 \\
 -i \\
 1 \\
\end{array}
\right),\quad
\lambda_2 =  \left(
\begin{array}{c}
 0 \\
 0 \\
 i \\
 1 \\
\end{array}
\right),\quad
\lambda_3=\left(
\begin{array}{c}
 -i \\
 1 \\
 0 \\
 0 \\
\end{array}
\right),\quad
\lambda_4=\left(
\begin{array}{c}
 i \\
 1 \\
 0 \\
 0 \\
\end{array}
\right)\label{eq:lambda}
\end{equation}
The first and the third eigenspinors correspond to eigenvalue $+1$ of $A$, and the other  two to eigenvalue  $-1$ of $A$.
We define these as the `rest spinors' $\lambda_i(0)$.
By acting the boost operator\footnote{We work in Weyl representation.}
\begin{equation}
\kappa = \sqrt{\frac{E+m}{2 m}}\left[\begin{array}{cc} 
\I + \frac{\s\cdot\p}{E+m} & \0 \\
\0 &  \I- \frac{\s\cdot\p}{E+m}
\end{array}\right]
\end{equation}
on these spinors we obtain the four eigenspinors for an arbitrary momentum $\lambda_i(\p) = \kappa\lambda_i(0)$. We implement our programme by solving the following four 
equations for $\tau_{ij}\in \R$:
\begin{align}
&m^{-1} \gamma_\mu p^\mu \lambda_1(\p) 
- \tau_{13}  \lambda_3(\p) =0,\quad
m^{-1} \gamma_\mu p^\mu \lambda_2(\p) 
- \tau_{24}\tau  \lambda_4(\p) =0\\
& m^{-1} \gamma_\mu p^\mu \lambda_3(\p) 
- \tau_{31}  \lambda_1(\p) =0,\quad
m^{-1} \gamma_\mu p^\mu \lambda_4(\p) 
- \tau_{42}  \lambda_2(\p) =0
\end{align}
and find that a single $\tau$, equal to unity, solves all the four equations and assures that while $\lambda_i(\p)$ do not  satisfy the Dirac equation they certainly satisfy the spinorial Klein-Gordon equation. We thus pass the first test for the viability of the theory to be Lorentz covariant. 
\vspace{5pt}

To study the CPT properties of $\lambda(\p)$ we 
introduce $\Theta$ (Wigner time reversal operator) and $\gamma$ 
\begin{align}
&\Theta =\left(\begin{array}{cc}
0 &-1\\
1 &0
\end{array}
\right),\quad
 \gamma = \frac{i}{4!} \epsilon_{\mu\nu\lambda\sigma}
\gamma^\mu\gamma^\nu\gamma^\lambda\gamma^\sigma = \left(\begin{array}{cc}
\I &\0\\
\0 & -\I
\end{array}\right)\label{eq:gamma5}
\end{align}
where $\epsilon_{\mu\nu\lambda\sigma}$ is the
completely antisymmetric 4th rank tensor with $\epsilon_{0123} = +1$.
The charge conjugation $\mathcal{C}$, parity $\mathcal{P}$, and time reversal $\mathcal{T}$, operators can then be written as
\begin{equation}
\mathcal{C} = \left(\begin{array}{cc}
\0 & i\Theta \\
-i\Theta &\0
\end{array}
\right) K,\quad
\mathcal{P}= m^{-1} \gamma_\mu p^\mu,\quad
\mathcal{T} = i \gamma \mathcal{C}
\end{equation}
where $K$ complex conjugates to the right. We then readily obtain
\begin{align}
&\mathcal{C} \lambda_1(\p) = - \lambda_4(\p),\quad
\mathcal{C} \lambda_2(\p) =  \lambda_3(\p),\quad
\mathcal{C} \lambda_3(\p) =  \lambda_2(\p),\quad
\mathcal{C} \lambda_4(\p) = - \lambda_1(\p), \\
&\mathcal{P} \lambda_1(\p) = \lambda_3(\p),\quad
\mathcal{P} \lambda_2(\p) =  \lambda_4(\p),\quad
\mathcal{P} \lambda_3(\p) =  \lambda_1(\p),\quad
\mathcal{P} \lambda_4(\p) =  \lambda_2(\p), \\
&\mathcal{T} \lambda_1(\p) = - i \lambda_4(\p),\quad
\mathcal{T} \lambda_2(\p) =  i \lambda_3(\p),\quad
\mathcal{T} \lambda_3(\p) =  -i \lambda_2(\p),\quad
\mathcal{T} \lambda_4(\p) = i \lambda_1(\p)
\end{align}
with the consequence that $(\mathcal{CPT})^2=\I$, with $\mathcal{C}^2=\I$, $ \mathcal{P}^2=\I$, $\mathcal{T}^2=-\I$. And, 
$
\{\mathcal{C},\mathcal{P}\} = 0.
$

\vspace{5pt}
As in the case for Elko~\cite{Ahluwalia:2019dv}, here too we find that under the Dirac dual the $\lambda(\p)$  have null norm. As such we define a new dual: 
\begin{equation}
\gdualn{\lambda}_1(\p) =\overline{\lambda}_3(\p),\quad
\gdualn{\lambda}_2(\p) =\overline{\lambda}_4(\p),\quad
\gdualn{\lambda}_3(\p) =\overline{\lambda}_1(\p),\quad
\gdualn{\lambda}_4(\p) =\overline{\lambda}_2(\p)
\end{equation}
 After re-norming the rest eigenspinors by a multiplicative factor of $\sqrt{m}$,  the new dual gives the following Lorentz invariant orthonormality relations 
 \begin{align}
&\gdualn{\lambda}_i(\p) {\lambda}_i(\p) = +2m,\quad i=1,2,\\
&\gdualn{\lambda}_i(\p) {\lambda}_i(\p) = + 2m,\quad i=3,4
\end{align}
and the `spin sums'
\begin{align}
\sum_{i=1,2}{\lambda}_i(\p) \gdualn{\lambda}_i(\p)  = 2 m\left(
\begin{array}{cccc}
0 & 0 & 0 & 0\\
0 & 0 & 0 & 0\\
0 & 0 & 1 & 0\\
0 & 0 & 0 &1
\end{array}\right),\quad
\sum_{i=3,4}{\lambda}_i(\p) \gdualn{\lambda}_i(\p)  = 2 m\left(
\begin{array}{cccc}
1 & 0 & 0 & 0\\
0 & 1 & 0 & 0\\
0 & 0 & 0 & 0\\
0 & 0 & 0 &0
\end{array}\right) \label{eq:ss}
\end{align}
leading to the completeness relation
\begin{equation}
\sum_{i=1,2} {\lambda}_i(\p) \gdualn{\lambda}_i(\p) +
\sum_{i=3,4} {\lambda}_i(\p) \gdualn{\lambda}_i(\p) =2 m \I \label{eq:spinsums}
\end{equation}
The appearance of the plus, rather than minus, sign between the two terms above would eventually justify the title of this communication.
We thus introduce a new spin one half quantum field with  $\lambda_i(\p)$ as its expansion co-efficients:
\begin{equation}
\mathfrak{b}(x) \stackrel{\mathrm{def}}{=}
\int\frac{\mbox{d}^3 p}{(2\pi)^3}
\frac{1}{\sqrt{2 m E(\p)}}
\bigg[
\sum_{i=1,2} {a}_i(\p)\lambda_i(\p)\big] e^{-i p\cdot x}+ \sum_{i=3,4} b^\dagger_i(\p)\lambda_i(\p)\big] e^{i p\cdot x}\bigg]
\end{equation}
with 
\begin{equation}
\gdualn{\mathfrak{b}}(x) \stackrel{\mathrm{def}}{=}
\int\frac{\mbox{d}^3 p}{(2\pi)^3}
\frac{1}{\sqrt{2 m E(\p)}}
\bigg[
\sum_{i=1,2} a^\dagger_i(\p)\gdualn{\lambda}_i(\p)\big] e^{i p\cdot x}+ \sum_{i=3,4} b_i(\p)\gdualn{\lambda}_i(\p)\big] e^{-i p\cdot x}\bigg]
\end{equation}
as its adjoint.
At this stage we do not fix the statistics to be fermionic 
\begin{equation}
\left\{a_i(\p),a^\dagger_j(\p)\right\} =(2 \pi)^3 \delta^3(\p-\p^\prime)\delta_{ij}, \quad \left\{a_i(\p), a_j(\p^\prime)\right\} = 0 =
 \left\{a^\dagger_i(\p), a^\dagger_j(\p^\prime)\right\}\label{eq:b}
\end{equation}
or bosonic
\begin{equation}
\left[a_i(\p),a^\dagger_j(\p)\right] =(2 \pi)^3 \delta^3(\p-\p^\prime)\delta_{ij}, \quad \left[a_i(\p), a_j(\p^\prime)\right] = 0 =
 \left[a^\dagger_i(\p), a^\dagger_j(\p^\prime)\right]\label{eq:bd}
\end{equation}
and assume similar anti-commutation, or commutation, relations for $b_i(\p)$ and $b_i^\dagger(\p)$. 
\vspace{5pt}

 To determine the statistics for the 
$\mathfrak{b}(x)$ and $\gdualn{b}(x)$ system we consider two events, $x$ and $x'$, and note that  the amplitude to propagate from $x$ to $x'$ is then
\begin{align}
\mathcal{A}_{x\to x^\prime}   =  \xi \Big(\underbrace{\langle\hspace{3pt}\vert
\mathfrak{b}(x^\prime)\gdualn{\mathfrak{b}}(x)\vert\hspace{3pt}\rangle \theta(t^\prime-t)
\pm  \langle\hspace{3pt}\vert
\gdualn{\mathfrak{b}}(x) \mathfrak{b}(x^\prime)\vert\hspace{3pt}\rangle \theta(t-t^\prime)}_{\langle\hspace{4pt}\vert \mathfrak{T} ( \mathfrak{b}(x^\prime) \gdualn{\mathfrak{b}}(x)\vert\hspace{4pt}\rangle}\Big)\label{eq:Axtoxprime}
\end{align}
where 
\begin{itemize}
 \item[\textemdash] the plus sign holds for the bosons and the minus sign for the fermions,

\item[\textemdash] $\xi\in\C$ is to be determined from the normalisation condition that
$
 \mathcal{A}_{x\to x^\prime} 
$
integrated over all possible separations $x-x^\prime$
 be unity (or, more precisely  $e^{i\gamma}$, with $\gamma\in \R$). 
\item[\textemdash] 
and $\mathfrak{T}$ is the time ordering operator.
\end{itemize}
The two vacuum expectation values that appear in $\mathcal{A}_{x\to x^\prime} $ evaluate to the following expressions
 \begin{align}
\langle\hspace{3pt}\vert
\mathfrak{b}(x^\prime)\gdualn{\mathfrak{b}}(x)\vert\hspace{3pt}\rangle  & =\int\frac{\text{d}^3p}{(2 \pi)^3}\left(\frac{1}{2 m E(\p)}\right)
 e^{-ip\cdot(x^\prime-x)}
 \sum_{i=1,2}\lambda_i(\p)
\gdualn\lambda_i(\p) \label{eq:amplitudeP-newS}
\\
\langle\hspace{3pt}\vert
\gdualn{\mathfrak{b}}(x) \mathfrak{b}(x^\prime)\vert\hspace{3pt}\rangle 
  & =  \int\frac{\text{d}^3p}{(2 \pi)^3}\left(\frac{1}{2 m E(\p)}\right)
 e^{ip\cdot(x^\prime-x)}
 \sum_{i=3,4}\lambda_i(\p)
\gdualn\lambda_i(\p) .\label{eq:amplitudeP-newA}
\end{align}
The two Heaviside step functions of equation
(\ref{eq:Axtoxprime}) can now be replaced by their integral representations
\begin{align}
\theta(t^\prime-t) &= \lim_{\epsilon\to 0^+} \int\frac{\text{d}\omega}{2\pi i}
\frac{e^{i \omega (t^\prime-t)}}{\omega- i \epsilon} \\
\theta(t-t^\prime) &= \lim_{\epsilon\to 0^+} \int\frac{\text{d}\omega}{2\pi i}
\frac{e^{i \omega (t-t^\prime)}}{\omega- i \epsilon}
\end{align}
where $\epsilon,\omega\in\R$. Using these results, and  
\begin{itemize}
\item  substituting $\omega \to p_0 = -\omega+E(\p)$ in the first term and  $\omega \to p_0 = \omega- E(\p)$ in the second term
\item and using the results (\ref{eq:spinsums})
for the spin sums
\end{itemize}
we are forced -- by internal consistency of the resulting formalism -- to pick the plus sign in (\ref{eq:Axtoxprime}), giving 
\begin{equation}
\mathcal{A}_{x\to x^\prime}  = i \,2 \xi \int\frac{\text{d}^4 p}{(2 \pi)^4}\,
e^{-i p_\mu(x^{\prime\mu}-x^\mu)}
\frac{\I}{p_\mu p^\mu -m^2 + i\epsilon}\label{eq:AmplitudeWithXi}
\end{equation}
This is equivalent to the choice (\ref{eq:bd}) over (\ref{eq:b}). The normalisation $\xi$ is seen to be \cite{Ahluwalia:2019dv}
\begin{equation}
\xi = \frac{i m^2}{2}
\end{equation}
resulting in
\begin{equation}
\mathcal{A}_{x\to x^\prime}  = - m^2 \int\frac{\text{d}^4 p}{(2 \pi)^4}\,
e^{-i p_\mu(x^{\prime\mu}-x^\mu)}
\frac{\I}{p_\mu p^\mu -m^2 + i\epsilon}
\end{equation}
We define the Feynman-Dyson propagator 
\begin{align}
S_{\textrm{FD}}(x^\prime-x) & \stackrel{\textrm{def}}{=} - \frac{1}{m^2} 
\mathcal{A}_{x\to x^\prime}\nonumber\\
 &= \int\frac{\text{d}^4 p}{(2 \pi)^4}\,
e^{-i p_\mu(x^{\prime\mu}-x^\mu)}
\frac{\I_4}{p_\mu p^\mu -m^2 + i\epsilon}\label{eq:FD-prop-b}
\end{align}
so that
\begin{equation}
\left(\partial_{\mu^\prime} \partial^{\mu^\prime} \I_4 + m^2\I_4\right)
S_{\textrm{FD}}(x^\prime-x)  = -  \delta^4(x^\prime - x)\label{eq:KGDiracDelta}
\end{equation}
In terms of the new field $\mathfrak{b}(x)$ and its adjoint $\gdualn{\mathfrak{b}}(x)$ it takes the form
\begin{equation}
S_{\textrm{FD}}(x^\prime-x)= -\frac{i}{2}\left\langle\hspace{4pt}\left\vert \mathfrak{T} ( \mathfrak{b}(x^\prime) \gdualn{b}(x)\right\vert\hspace{4pt}\right\rangle
\end{equation}
and establishes the mass dimension of the field to be one, leading to the following free field Lagrangian density
\begin{equation}
\mathfrak{L}_0(x) =\partial^\mu\gdualn{\mathfrak{b}}\,\partial_\mu {\mathfrak{b}}(x) - m^2 \gdualn{\mathfrak{b}}(x) \mathfrak{b}(x)\label{eq:fieldlagrangian}
\end{equation}
This leads to the momentum conjugate to $\mathfrak{b}(x)$
\begin{equation}
\mathfrak{p}(x) = \frac{\partial \mathfrak{L}_0(x)}
{\partial {\dot{\mathfrak{b}}(x)}} = \frac{\partial}{\partial t}\gdualn{\mathfrak{b}}(x).
\end{equation}
Using the spin sums given in equation (\ref{eq:ss}) we determine the locality structure of the new bosonic field to be
\begin{align}
&\left[\mathfrak{b}(t,\x),\;\mathfrak{p}(t,\x^\prime) \right] = i \delta^3\left(\x-\x^\prime\right) \openone_\ell,\quad\\
&\left[ \mathfrak{b}(t,\x),\;\mathfrak{b}(t,\x^\prime) \right]= 0, 
\left[ \mathfrak{p}(t,\x),\;\mathfrak{p}(t,\x^\prime) \right] = 0.\label{eq:lac-2and3}
\end{align}
where
\begin{equation}
\openone_\ell \stackrel{\mathrm{def}}{=} \left(
\begin{array}{cccc}
-1 & 0 &  0 & 0\\
0 & -1 & 0 & 0\\
0 & 0 & 1 & 0\\
0 & 0& 0& 1
\end{array}\right)
\end{equation}
Had we taken the locality phases\footnote{The constant factors mentioned just above equation~(\ref{eq:lambda}).}  to be $-1$ for of the $\lambda_3$ and 
$\lambda_4$, the entire formalism would have run precisely as above with the difference that the locality structure of the new bosonic field would have changed to 
\begin{align}
&\left[\mathfrak{b}(t,\x),\;\mathfrak{p}(t,\x^\prime) \right] = i \delta^3\left(\x-\x^\prime\right) \openone_r,\quad\\
&\left[ \mathfrak{b}(t,\x),\;\mathfrak{b}(t,\x^\prime) \right]= 0, 
\left[ \mathfrak{p}(t,\x),\;\mathfrak{p}(t,\x^\prime) \right] = 0.\label{eq:lac-2and3}
\end{align}
where
\begin{equation}
\openone_r \stackrel{\mathrm{def}}{=} \left(
\begin{array}{cccc}
1 & 0 &  0 & 0\\
0 & 1 & 0 & 0\\
0 & 0 & -1 & 0\\
0 & 0& 0& -1
\end{array}\right)
\end{equation}
We thus conclude that in order to preserve locality parity must be maximally violated with the left and right projections of $\mathfrak{b}(x)$ serving as two independent local fields. This makes us suspect that neutrinos may be bosonic rather than fermionic as was first speculated by Dolgov and Smirnov~\cite{Dolgov:2005qi}.

\vspace{21pt}
\noindent
\textbf{Acknowledgement}\textemdash The author wishes to thank Suresh Chand for discussions.

\newpage
\appendix
\section{Square roots of $4\times 4$ identity matrix}\label{AppA}

\noindent
\underline{Antisymmetric solutions}, up to a sign
\begin{equation}
A=\left(
\begin{array}{cccc}
0 & -i & 0 & 0\\
i & 0 & 0 & 0\\
0 & 0 & 0 & -i\\
0 & 0& i & 0
\end{array}
\right),\quad
A=\left(
\begin{array}{cccc}
0 & -i & 0 & 0\\
i & 0 & 0 & 0\\
0 & 0 & 0 & i\\
0 & 0& -i & 0
\end{array}
\right)\label{eq:a1}
\end{equation}
With the definitions
\begin{align}
& \epsilon \stackrel{\mathrm{def}}{=} \alpha + i \beta\\
& \varepsilon\stackrel{\mathrm{def}}{=}  i\sqrt{(\alpha+ i \beta)^2+1}  =  i \sqrt{\epsilon^2+1}\\
& \eta \stackrel{\mathrm{def}}{=} \gamma + i \delta\\
& \zeta \stackrel{\mathrm{def}}{=} i\sqrt{(\gamma+ i \delta)^2+1}=
i\sqrt{\eta^2+1}\\
& \xi \stackrel{\mathrm{def}}{=} i\sqrt{(\gamma+i\delta)^2+(\alpha+i\beta)^2+1} = i\sqrt{\eta^2+\epsilon^2+1}
\end{align}
we have the following twelve additional $A$'s (again, up to a sign)
\begin{align}
&A=\left(
\begin{array}{cccc}
0 & \varepsilon& 0 & \epsilon\\
-\varepsilon & 0 & -\epsilon & 0\\
0 & \epsilon & 0 & -\varepsilon\\
-\epsilon & 0& \varepsilon & 0
\end{array}
\right),\quad
A=\left(
\begin{array}{cccc}
0 & -\varepsilon & 0 & \epsilon\\
\varepsilon & 0 & -\epsilon & 0\\
0 & \epsilon& 0 & \varepsilon\\
-\epsilon & 0& -\varepsilon & 0
\end{array}
\right)\\
& A=\left(
\begin{array}{cccc}
0 & \varepsilon& 0 & \epsilon\\
-\varepsilon & 0 & \epsilon & 0\\
0 & -\epsilon & 0 & \varepsilon\\
-\epsilon & 0& -\varepsilon & 0
\end{array}
\right),\quad
 A=\left(
\begin{array}{cccc}
0 & -\varepsilon& 0 & \epsilon\\
\varepsilon & 0 & \epsilon & 0\\
0 & -\epsilon & 0 &-\varepsilon\\
-\epsilon & 0& \varepsilon & 0
\end{array}
\right)
\end{align}
\begin{align}
& A=\left(
\begin{array}{cccc}
0 & \zeta& \eta & 0\\
-\zeta & 0 & 0 & -\eta\\
 -\eta &0 & 0 &\zeta\\
0 & \eta& -\zeta & 0
\end{array}
\right),\quad
A=\left(
\begin{array}{cccc}
0 &  -\zeta& \eta & 0\\
\zeta & 0 & 0 & -\eta\\
 -\eta &0 & 0 &-\zeta\\
0 & \eta& \zeta & 0
\end{array}
\right)\\
& A=\left(
\begin{array}{cccc}
0 & \zeta& \eta & 0\\
- \zeta & 0 & 0 & \eta\\
 -\eta &0 & 0 &-\zeta\\
0 & -\eta  & \zeta & 0
\end{array}
\right),\quad 
A=\left(
\begin{array}{cccc}
0 & -\zeta& \eta & 0\\
\zeta & 0 & 0 & \eta\\
 -\eta &0 & 0 &\zeta\\
0 & -\eta& -\zeta & 0
\end{array}
\right)
\end{align}
\begin{align}
&A=\left(
\begin{array}{cccc}
0 & \xi& \eta & \epsilon\\
-\xi & 0 & -\epsilon & \eta\\
 -\eta &\epsilon & 0 &-\xi\\
-\epsilon & -\eta& \xi & 0
\end{array}
\right),\quad
 A=\left(
\begin{array}{cccc}
0 & -\xi& \eta & \epsilon\\
\xi & 0 & -\epsilon & \eta\\
 -\eta &\epsilon & 0 &\xi\\
-\epsilon & -\eta& -\xi & 0
\end{array}
\right),\\
& A=\left(
\begin{array}{cccc}
0 & \xi& \eta & \epsilon\\
-\xi & 0 & \epsilon & -\eta\\
 -\eta &-\epsilon & 0 &\xi\\
-\epsilon & \eta& -\xi & 0
\end{array}
\right),\quad
A=\left(
\begin{array}{cccc}
0 & -\xi& \eta & \epsilon\\
\xi & 0 & \epsilon & -\eta\\
 -\eta &-\epsilon & 0 &-\xi\\
-\epsilon & \eta& \xi & 0
\end{array}
\right)
\end{align}
\vspace{7pt}
\underline{Symmetric solutions (restricted to $\mbox{diag}\{0,0,0,0\}$)}, up to a sign:\footnote{The restriction to $\mbox{diag}\{0,0,0,0\}$ is only for the ease of calculations. Relaxing this restriction gives a large number of additional solutions. With some insightful strategy, with say Mathematica, the reader can easily examine those additional roots.
}
\begin{equation}
A=\left(
\begin{array}{cccc}
0 & -1 & 0 & 0\\
-1 & 0 & 0 & 0\\
0 & 0 & 0 & -1\\
0 & 0& -1 & 0
\end{array}
\right),\quad
A=\left(
\begin{array}{cccc}
0 & -1 & 0 & 0\\
-1 & 0 & 0 & 0\\
0 & 0 & 0 & 1\\
0 & 0& 1 & 0
\end{array}
\right)
\end{equation}
Introducing 
\begin{align}
& \rho \stackrel{\mathrm{def}}{=} i\sqrt{(\alpha + i\beta-1)(\alpha + i\beta+1)}\\
& \sigma \stackrel{\mathrm{def}}{=} i\sqrt{(\gamma + i\delta-1)(\gamma+ i\delta+1)}
\end{align}
we have the following twelve additional solutions (again, up to a sign)
\begin{align}
&A =\left(\begin{array}{cccc}
0 & 0 & \rho &\epsilon\\
0 & 0& -\epsilon &\rho\\
\rho & -\epsilon &0 & 0\\
\epsilon & \rho & 0 & 0
\end{array}
\right),\quad
A =\left(\begin{array}{cccc}
0 & 0 & -\rho &\epsilon\\
0 & 0& -\epsilon &-\rho\\
-\rho & -\epsilon &0 & 0\\
\epsilon & -\rho & 0 & 0
\end{array}
\right)\\
&A = \left(\begin{array}{cccc}
0 & \rho & 0 & \epsilon\\
\rho & 0 & -\epsilon &0\\
0 & -\epsilon & 0 & \rho\\
\epsilon & 0 & \rho & 0
\end{array}
\right),\quad
A = \left(\begin{array}{cccc}
0 & -\rho & 0 & \epsilon\\
-\rho & 0 & -\epsilon &0\\
0 & -\epsilon & 0 & -\rho\\
\epsilon & 0 &- \rho & 0
\end{array}
\right)
\end{align}
\begin{align}
& A = \left(\begin{array}{cccc}
0 & 0 & \rho & \epsilon \\
0 & 0 & \epsilon & -\rho\\
\rho & \epsilon &  0 & 0\\
\epsilon & -\rho & 0 & 0
\end{array}
\right),\quad
A =  \left(\begin{array}{cccc}
0 & 0 & -\rho & \epsilon\\
0 & 0 & \epsilon & \rho \\
-\rho & \epsilon & 0 & 0\\
\epsilon & \rho & 0 & 0
\end{array}
\right)
\\
&A = \left(\begin{array}{cccc}
0 & \rho &0 & \epsilon\\
\rho & 0 & \epsilon & 0\\
0 & \epsilon & 0 & -\rho\\
\epsilon & 0 & -\rho  & 0
\end{array}
\right),\quad
A =  \left(\begin{array}{cccc}
0 & -\rho & 0 &\epsilon \\
-\rho & 0 & \epsilon & 0\\
0 & \epsilon & 0 &\rho\\
\epsilon & 0 & \rho & 0
\end{array}
\right) 
\end{align}
\begin{align}
& A =\left(\begin{array}{cccc}
0 & \sigma &  \eta & 0\\
\sigma & 0 & 0 &-\eta\\
\eta & 0 & 0 &\sigma\\
0 & -\eta & \sigma & 0
\end{array}
\right),\quad
 A =\left(\begin{array}{cccc}
0 & -\sigma &  \eta & 0\\
- \sigma & 0 & 0 &-\eta\\
\eta & 0 & 0 &-\sigma\\
0 & -\eta & -\sigma & 0
\end{array}
\right)\\
& A =\left(\begin{array}{cccc}
0 & \sigma &  \eta & 0\\
\sigma & 0 & 0 &\eta\\
\eta & 0 & 0 &-\sigma\\
0 & \eta & -\sigma & 0
\end{array}
\right),\quad 
A =\left(\begin{array}{cccc}
0 & -\sigma &  \eta & 0\\
-\sigma & 0 & 0 &\eta\\
\eta & 0 & 0 &\sigma\\
0 & \eta & \sigma & 0
\end{array}
\right)
\end{align}

\noindent
\textbf{References}

\begin{thebibliography}{1}
\expandafter\ifx\csname url\endcsname\relax
  \def\url#1{\texttt{#1}}\fi
\expandafter\ifx\csname urlprefix\endcsname\relax\def\urlprefix{URL }\fi
\expandafter\ifx\csname href\endcsname\relax
  \def\href#1#2{#2} \def\path#1{#1}\fi

\bibitem{Dirac:1928hu}
P.~A.~M. Dirac, {The quantum theory of the electron}, Proc. Roy. Soc. Lond.
  A117 (1928) 610--624.

\bibitem{Weinberg:1995mt}
S.~Weinberg, {The quantum theory of fields. Vol. 1: Foundations}, Cambridge
  University Press, 2005.

\bibitem{Majorana:1937vz}
E.~Majorana, {Theory of the symmetry of electrons and positrons}, Nuovo Cim. 14
  (1937) 171--184.

\bibitem{Ahluwalia:2019dv}
D.~Ahluwalia, {Mass Dimension One Fermions (Cambridge monographs on
  mathematical physics)}, Cambridge University Press, 2019.

\bibitem{Dolgov:2005qi}
A.~D. Dolgov, A.~{\relax Yu}. Smirnov, {Possible violation of the
  spin-statistics relation for neutrinos: Cosmological and astrophysical
  consequences}, Phys. Lett. B621 (2005) 1--10.

\end{thebibliography}

\end{document}